\documentclass[preprint,12pt]{elsarticle}

\usepackage{amssymb}
\usepackage{amsmath}

\usepackage{xcolor}
\usepackage[normalem]{ulem}

\usepackage{hyperref}
\usepackage{tikz}
\usepackage{longtable}
\usepackage{booktabs}
\usepackage{chngcntr}

\journal{Solid State Sciences}

\begin{document}

\begin{frontmatter}


\title{The seeding method: A test case for classical nucleation theory in small systems}

\author[label1]{Thomas Philippe}
\author[label1]{Yijian Wu}
\author[label1]{Aymane Graini}

\affiliation[label1]{organization={Laboratoire de Physique de la Matière Condensée, Ecole polytechnique, CNRS, IP Paris},
            city={Palaiseau},
            postcode={91128}, 
            country={France}}

\begin{abstract}
Molecular dynamics simulations are widely used to investigate nucleation in first-order phase transitions. Brute-force simulations, though popular, are limited to conditions of high metastability, where the critical cluster and the nucleation barrier are small. The seeding method has recently emerged as a powerful alternative for exploring lower supersaturation regimes by initiating simulations with a pre-formed nucleus. In confined systems (NVT ensemble), the seeded simulations are particularly effective for determining stable cluster properties and provide a stringent test case for classical nucleation theory (CNT). In this work, we perform NVT seeded simulations of Lennard-Jones condensation in small systems and compare them with CNT predictions based on several thermodynamic models, including equations of state, perturbation theory, and ideal gas approximation. We find that CNT accurately predicts stable cluster radii across a wide range of conditions. Notably, even the very simple ideal gas approximation proves useful for initializing seeded simulations. Furthermore, seeded simulation results correspond to the critical cluster radii of infinite systems: CNT predictions with good equations of state show very good agreement with simulations, while the perturbation theory and the ideal gas approximation perform well at low temperatures but deviate significantly at high temperatures. 

\end{abstract}


\begin{highlights}
\item We perform NVT seeded simulations of Lennard-Jones condensation in small systems to stabilize critical droplets and compare their properties with classical nucleation theory predictions.
\item We demonstrate how the classical nucleation theory can guide the setup of NVT seeded simulations and concomitantly evaluate the existing thermodynamic models.
\end{highlights}


\end{frontmatter}


\section{Introduction}
\label{sec:intro}
\pagenumbering{arabic}
\normalsize Nucleation is a rare and thermally activated process. In molecular dynamics (MD) simulations, only small critical clusters with low nucleation barrier --- in the regime of high supersaturation --- are accessible with brute-force approaches~\cite{YM_method_JCP1998, auerPredictionAbsoluteCrystalnucleation2001, Reguera_JCP2009, Tanaka_JCP2005, Zhukhovitskii_JCP2016, MD_LJ_JCP2000, Wedekin_Reguera_JCP2009}, even when the simulation system is large~\cite{Diemand_JCP2013, Diemand_JCP2014, Diemand_PRE2014}. While nucleation rates can be readily measured from brute-force simulations~\cite{Hangi-Talkner1990, YM_method_JCP1998}, identifying critical clusters remains highly challenging~\cite{Angelil_JCP2014}. A common strategy to estimate the critical cluster size is to apply the nucleation theorem~\cite{Kashchiev_JCP1982} from measured nucleation rate~\cite{Diemand_JCP2013}. However, in the high-supersaturation regime, classical nucleation assumptions become questionable~\cite{Wedekin_Reguera_JCP2009}. 
To overcome these limitations, a variety of rare-event simulation techniques have been developed to promote the formation of critical cluster~\cite{Izrailev1999, Dellago2002Jul, Bolhuis2002Oct, Peters2006Aug, Allen2009Oct,  Kastner2011Nov, Hartmann2013Dec, Patel2014Feb, Comer2015Jan, Bernardi2015May, Mullen2015Jun, Bussi2020Apr, Raucci2022Feb, Burgin2023Jan,Vega_JCP_2024,Ronneberger_2015,Finney_2024}. In parallel, another approach has emerged: the seeding technique \cite{Vega_Seeding_JCP2016, Seeding_NVT_Cavitation_PRE2020, Vega-Seeding-NVT_PRE2020, Vega_PRE2019}. The technique was employed to investigate cavitation and condensation, as well as crystal nucleation \cite{Vega_Seeding_JCP2016,Sun_PRL2018}. In seeded MD simulations, the system is initialized with a pre-formed nucleus of the new phase. At constant pressure and temperature (NPT ensemble), the inserted seed may either grow (if post-critical) or redissolve (if pre-critical). At the exact critical size, the cluster corresponds to an unstable equilibrium, with a $50\%$ probability of growth or dissolution. Once the critical cluster is identified, classical nucleation theory (CNT)~\cite{Kashchiev_JCP1982, becker_kinetische_1935, volmer_kinetik_1939, zeldovich_theory_1942, frenkel_kinetic_1946, Turnbull_1949, feder_homogeneous_1966}, on which the approach relies, can then be applied to estimate the nucleation rate~\cite{Vega_Seeding_JCP2016}. This approach is conceptually simple and intuitive, but in practice requires many independent simulations to pinpoint the condition at which the seed is critical. Moreover, because the seeds typically grow or redissolve rapidly, NPT seeding can be computationally intensive when aiming for robust statistics.

In order to make the seeding technique more versatile, the method was then extended to the NVT ensemble~\cite{Seeding_NVT_Cavitation_PRE2020, Vega-Seeding-NVT_PRE2020, Vega_hardspheres_JCP2020, Vega_JCP2022,Vega_JCP2023}. This approach builds on the framework of classical nucleation in small systems~\cite{Schmelzer_2006, Schmelzer_2007, Schmelzer_2011, Reguera_2003, Wilhelmsen_2014_communication, Wilhelmsen_2015, Philippe_PRE_2017}. Mass conservation, together with chemical and mechanical equilibrium conditions, give rise to two critical states: one unstable and the other stable~\cite{Reguera_2003}. The objective of NVT seeded simulations is to achieve the stable equilibrated configuration, which corresponds to the critical unstable cluster in the NPT ensemble or, equivalently, in the infinite system at the corresponding supersaturation~\cite{Vega_Seeding_JCP2016,Vega_JCP2023}. Yet, nucleation in confined systems is also known to present some peculiarities, such as superstabilization, and finite-size effects must be carefully accounted for when applying the NVT seeding approach. 

In this paper, we apply CNT to study the condensation of nanodroplets in the NVT ensemble. Lennard-Jones particles are chosen as a model system, and \textit{a priori} for testing some thermodynamic models, since many efforts have been devoted to developing, for instance, accurate equation of states (EOS) for this system \cite{johnson_lennard-jones_1993,kolafa_lennard-jones_1994,stephan_review_2020}. We demonstrate then how CNT can guide the setup of NVT seeded simulations and concomitantly evaluate the existing thermodynamic models. We first briefly describe the NVT seeding method for condensation of Lennard-Jones particles. 

\section{Simulation details}
\label{sec:simulation}
\normalsize Simulations were performed using the open-source package \texttt{LAMMPS} \cite{LAMMPS,NGUYEN2017113}. For the Lennard-Jones interactions, we used a large cutoff of $6.78\sigma$ with $\sigma$ the length parameter \cite{baidakov_self-diffusion_2011,baidakov_metastable_2007}. Lennard-Jones units are used in the following. Energies are scaled by $\epsilon$, which is the energy depth of the untruncated Lennard-Jones potential. Lengths are scaled by $\sigma$, temperatures by $\epsilon/k_\mathrm{B}$, densities by $\sigma^{-3}$, and time by $\tau = \sqrt{m \sigma^2 / \epsilon}$ with $m$ the mass of the particle. The timestep employed is $0.005\tau$. Cubic periodic boundary conditions are used. For temperature control, we employed the Nose-Hoover thermostat \cite{Martyna_JCP1994,LAMMPS} implemented in \texttt{LAMMPS}. We use the following procedure for performing seeded simulations. The first step is to prepare the seed. The seed is obtained from a NVT run in a box of arbitrary size to equilibrate the single phase liquid state at the density of the phase diagram $\bar{\rho}_l$, which sets the number of particles in the chosen volume. Then a sphere of a given size $R$ is extracted from the liquid phase and inserted in a cubic box of size $L$. This sets the initial number of liquid particles, $N_l$, in the seeded simulation:
\begin{equation}
    N_l=\frac{4}{3}\pi R^3 \bar{\rho}_l.
\end{equation}
It remains to choose the initial number of vapor particles, $N_v$, which are randomly distributed outside the liquid droplet. The box density, $\rho$, is then given by:
\begin{equation}
    \rho = \frac{N_l+N_v}{L^3}.
\end{equation}
Therefore, for a given box size $L$, there are two independent parameters, $R$ and $\rho$, that must be carefully chosen to guarantee that the subsequent NVT run leads to the stabilization of the liquid droplet to the expected (stable) equilibrium in the confined system. However, nucleation in small systems present some peculiarities, detailed in the following section, that can complicate the choice of the parameters of NVT seeded simulations. For instance, if a given triplet $(L,R,\rho)$ implies a too large value of the initial vapor density, spontaneous and undesired thermally activated nuclei will form in the vapor phase. The seed can also redissolve if the initial vapor density is too small. Moreover, for a given $\rho$, no stable cluster will exist if the box size $L$ is too small, due to the so-called superstabilization effect \cite{Reguera_2003}. As we will show in the following, the NVT-seeding approach can greatly benefit from insights of classical nucleation theory.

\section{Classical nucleation theory}
\label{sec:CNT}
\normalsize In an infinite system, a supersaturated vapor phase is initially metastable, i.e. the formation of the stable phase (liquid) is energetically favorable for sufficiently large clusters. Yet, in a confined (finite) system, both the size of the critical cluster and the nucleation barrier are known to increase as compared with nucleation in the infinite system. Therefore, the phase transition is systemically delayed. But as the size of the confined system decreases, nucleation can be impeded and the initial state is no more metastable but becomes stable, because of mass conservation \cite{Reguera_2003}. Thus, for a given supersaturation this superstabilization effect can prevent nucleation to occur in too small systems. In larger systems however, generally, a new minimum in energy appears, associated with the formation of a nucleus that corresponds to a stable and final state, as growth is stopped. The calculation of this stable equilibrium is reminded in this section. Following many theoretical works dedicated to first-order phase transitions in confined systems \cite{Schmelzer_2006,Schmelzer_2007,Schmelzer_2011,Reguera_2003,Wilhelmsen_2014_communication,Wilhelmsen_2015,Philippe_PRE_2017,Wilhelmsen_JCP2014_SGAtheory}, we consider CNT in the NVT ensemble within the capillary approximation. $N$ is the total number of particles in the system, $V=L^3$ its volume, and $T$ the fixed temperature. The system therefore consists of a spherical liquid nucleus of radius $R$ (with homogeneous density $\rho_l$) in the vapor phase of density $\rho_v$. In the capillary approximation, the liquid and vapor phases are separated by a sharp interface with constant surface tension $\gamma$.

In the NVT (canonical) ensemble, the appropriate thermodynamic potential to consider is the Helmholtz free energy, $F$. Since the cluster is spherical, the variation of the total Helmholtz free energy of the system is given by \cite{Reguera_2003}
\begin{equation}
\label{dF}
    \mathrm{d}F=\left(p_v - p_l + \frac{2\gamma}{R} \right) \mathrm{d}V_l +\left( \mu_l - \mu_v \right)\mathrm{d}N_l,
\end{equation}
where $p_l$ and $p_v$ denote the liquid and vapor pressures, respectively, and $\mu_l$ and $\mu_v$ denote the chemical potentials of the liquid and vapor phases, respectively. $V_l$ is the volume of the droplet of radius $R$, which contains $N_l$ particles. At equilibrium, $\mathrm{d}F=0$, leading to the equality of the chemical potentials of the liquid and vapor phases,
\begin{equation}
\label{eqmu}
    \mu_l^* = \mu_v^*,
\end{equation}
and the Laplace relation
\begin{equation}
\label{laplace}
    p_l^* - p_v^* = \frac{2\gamma}{R^*}.
\end{equation}
Properties marked with $*$ are those at equilibrium. In an infinite system, or when the chemical potential of the vapor phase is fixed and known (in the grand canonical ensemble for instance), supersaturation is fixed and the above system admits only one solution that corresponds to the unstable equilibrium and sets the properties of the critical cluster, namely its density $\rho^*$ and its size $R^*$. However, in the canonical ensemble, the total number of particles (i.e., mass) is conserved, the vapor phase density is not fixed but depends on the number of molecules constituting the droplet,
\begin{equation}
\label{masscons}
    \rho = \Phi \rho_l + \left(1-\Phi \right) \rho_v,     
\end{equation}
where $\Phi=V_l/V$ is the volume fraction occupied by the liquid. The critical droplet must also satisfy the above condition. The combination of mass conservation and the equilibrium conditions has now two solutions, one corresponding to the critical and unstable cluster, and the other corresponding to a larger and stable droplet in equilibrium with the vapor. The size of the unstable critical cluster is known to slightly increase in a confined system \cite{Wilhelmsen_2014_communication}, as compared with that in the infinite system, which delays the phase transition. An approximated expression for the critical droplet in small systems can be obtained from expending the pressures and densities at first order \cite{Wilhelmsen_2015}. The larger droplet corresponds to a stable equilibrium. The objective of seeded simulations in the NVT ensemble is to equilibrate this droplet in the vapor phase.

\begin{figure}[p]
\centering
\includegraphics[width=0.7\textwidth]{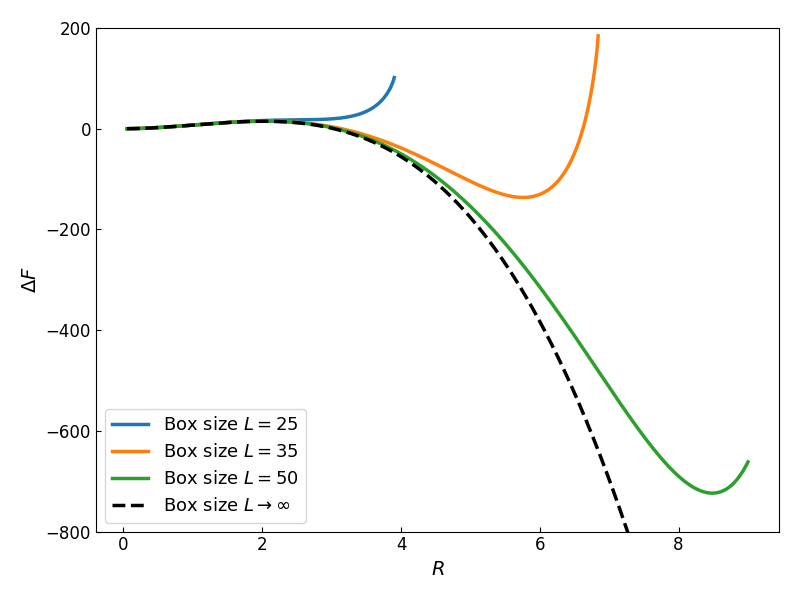}
\caption{Free energy associated with the formation of a spherical droplet of radius $R$ in a box of size $L$ with density $\rho=0.025$ at $T=0.8$.}
\label{fig:barrier}
\end{figure}

In practice, the critical densities of both the stable and unstable clusters are very close to that of the phase diagram, $\bar{\rho}_l$. If we further assume that the vapor phase is ideal and the liquid phase incompressible, one gets in reduced units the following relation
\begin{equation}
\label{exactR}
    \frac{2\gamma}{R^*} = \bar{\rho}_l T \ln{\frac{p_v^*}{\bar{p}_v}} - \left(p_v-\bar{p}_v \right),
\end{equation}
where we have used $\mu_l = \bar{\mu}_v + (p_l-\bar{p}_v)/\bar{\rho}_l$ with $\bar{p}_v$ the pressure of the saturated vapor and $\mu_v = \bar{\mu}_v + T \ln{(p_v/\bar{p}_v)}$ together with Eqs.\ref{eqmu} and \ref{laplace}. The last term in Eq.\ref{exactR} is usually very small and is commonly discarded and the well-known result is recovered for the critical size
\begin{equation}
\label{approxR}
    R^*=\frac{2\gamma}{\bar{\rho}_lT \ln{S^*}} ,
\end{equation}
with $S^*=p_v^*/\bar{p}_v$ the ratio between the vapor pressure at equilibrium and that of the saturated vapor. Since the vapor is assumed to be ideal gas, $S^*=\rho_v^*/\bar{\rho}_v$, where $\bar{\rho}_v$ is the saturated vapor density of the phase diagram. Together with Eq.~(\ref{masscons}) and $\rho_l^*=\bar{\rho}_l$, Eq.~(\ref{approxR}) constitutes the approximate problem derived by Reguera et al.~\cite{Reguera_2003} to find both the unstable and stable clusters in confined systems. However, if quantitative predictions are intended, a more rigorous thermodynamic description of the chemical potential should be required, as we shall see.

Aside from the coexistence conditions, it is always useful to examine the free energy landscape for nucleation. The free energy difference of the confined system, $\Delta F = F(R)-F(0)$, associated with the formation of a spherical droplet of radius $R$ in a box of size $L$ with density $\rho$, can be expressed as 
\begin{equation}
    \Delta F = - \bar{\rho}_l V_l T \ln{\frac{\rho_v}{\bar{\rho}_v}} + A_l \gamma + L^3T \left(\rho - \Phi \bar{\rho}_v + \rho_v \left(\Phi - 1 \right) +\rho \ln{\frac{\rho_v}{\rho}}\right).
\end{equation}
Again we assumed that the droplet is incompressible and the vapor is ideal. $A_l$ refers to the droplet surface. This free energy landscape is shown in Fig.~\ref{fig:barrier} for various box size with global density $\rho=0.025$ at temperature $T=0.8$. The value of $\gamma$ is taken from Ref.~\cite{baidakov_metastable_2007}. For an infinite system, the classical result is recovered: the free energy exhibits a single maximum, which corresponds to the critical size of the unstable droplet, obtained by setting $\rho_v = \rho$ and $\Phi = 0$. When the system is confined (for instance $L=50$), the free energy shows two extrema: the first is a maximum associated with the unstable state and the second is a minimum that gives the size of the stable droplet. When the system is too small, the free energy of the second extrema becomes larger than that of the initial state (as for $L=25$) and nucleation is impeded due to superstabilization. Evidently, a prior estimation of the stable equilibrium, if existing, will help in the choice of the initial droplet for designing efficient seeded simulations.

\section{Results}
\label{sec:results}
\begin{figure}[p]
\centering
\includegraphics[width=0.7\textwidth]{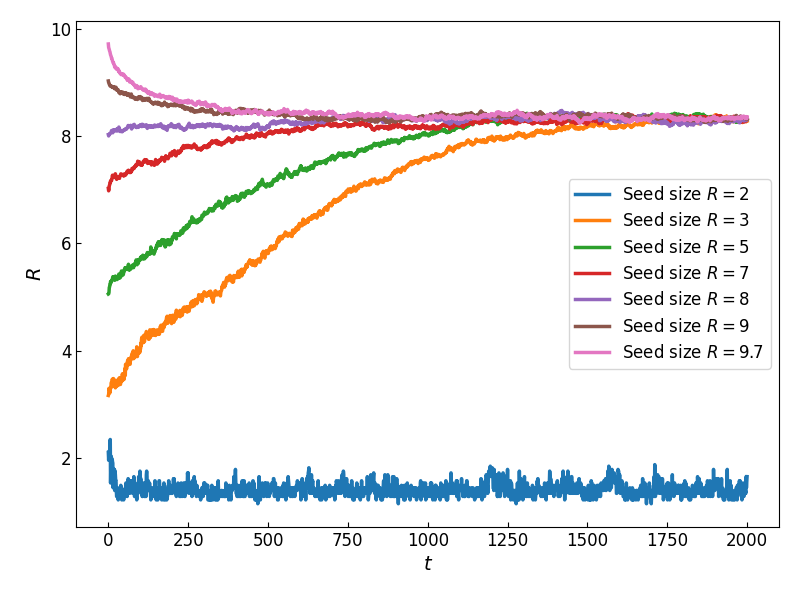}
\caption{Temporal evolution of the liquid droplet radius resulting from NVT seeded simulations in a box of size $L=50$ with global density $\rho=0.025$ at $T=0.8$. Various sizes for the liquid seed are employed to initialize the NVT simulations, ranging from $R=2$ to $R=9.7$. All seed sizes reach that of the stable droplet (near $R\sim 8.5$), except the smallest seed that redissolves.}
\label{fig:runs}
\end{figure}

We first examine the influence of the initial droplet in NVT seeded simulations for the condensation of Lennard-Jones particles. A liquid seed of density $\bar{\rho}_l$ is introduced with an initial radius $R$ in a box of size $L=50$ with global density $\rho=0.025$ at $T=0.8$. Mass conservation determines the vapor density of the initial state. Various sizes for the liquid seed are used, ranging from $R=2$ to $R=9.7$. The NVT simulations run for $2000\tau$. Fig.~\ref{fig:runs} reports the size of the liquid droplet with time. When the seed is supercritical ($R>3$), the stable equilibrium is reached. It corresponds to a droplet of size $R\sim8.5$ in equilibrium with the vapor phase. This stable configuration, which refers to a minimum in free energy, is in very good agreement with CNT predictions, see Fig.\ref{fig:barrier} for $L=50$. The smallest seed ($R=2$) is found to redissolve, which also aligns with CNT predictions (Fig.\ref{fig:barrier}). NVT runs suggest that the critical (unstable) radius is indeed comprised between $R=2$ and $R=3$ (Fig.\ref{fig:runs}). Note that, after disappearance of the smallest seed ($R=2$), thermal fluctuations produce the stochastic emergence of small clusters made of roughly $10-15$ particles, detected in Fig.\ref{fig:runs}. None of them are found to reach the critical size and enter the growth regime within the time window simulated in our NVT runs, as expected since supersaturation is low. These NVT seeded simulations demonstrate that the stable droplet is independent of the size of the initial seed, as long as it is chosen within the supercritical region ($R>3$ in this case). Nevertheless, the time for reaching this stable equilibrium is naturally size dependent, highlighting the benefit from initializing with a size close to the targeted equilibrium. In the following, CNT predictions from Eqs.~(\ref{eqmu}), (\ref{laplace}), and (\ref{masscons}) will be employed to initialize the seeding simulations.

We then investigate the condensation of Lennard-Jones particles over the temperature range from $0.7$ to $1.1$. The two solutions of Eqs.~(\ref{eqmu}), (\ref{laplace}), and (\ref{masscons}) are obtained using chemical potential computed from the EOS developed by Johnson, Zollweg, and Gubbins (JZG)~\cite{johnson_lennard-jones_1993}. The EOS are combined with a mean-field correction for the cut potential \cite{johnson_lennard-jones_1993}, using the same cutoff ($6.78$) as in the simulations. Within this temperature range, the phase diagram predicted by the JZG EOS is found to be in excellent agreement with that obtained from MD simulations of the two-phase system at equilibrium (not shown). The surface tension of the planar vapor/liquid interface at equilibrium is classically calculated from MD simulations using the Kirkwood pressure tensor method~\cite{Kirkwood_JCP_1950}. The results follow the expected power law \cite{baidakov_metastable_2007}, $\gamma = \gamma_0 \left( 1-T/T_c\right)^a$, with $\gamma_0=2.838$, $a=1.252$, and $T_c=1.305$ the critical temperature.

\begin{figure}[p]
\begin{tikzpicture}
\node (fig1) at (-3,0) {\includegraphics[width=0.45\textwidth]{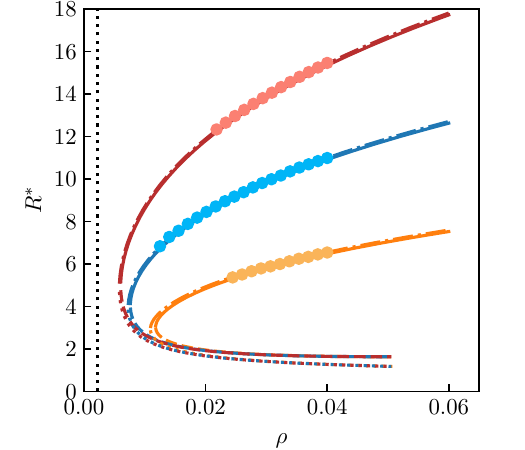}};
\node[] at (-6,2.7) {(a)};
\node (fig2) at (3,0) {\includegraphics[width=0.45\textwidth]{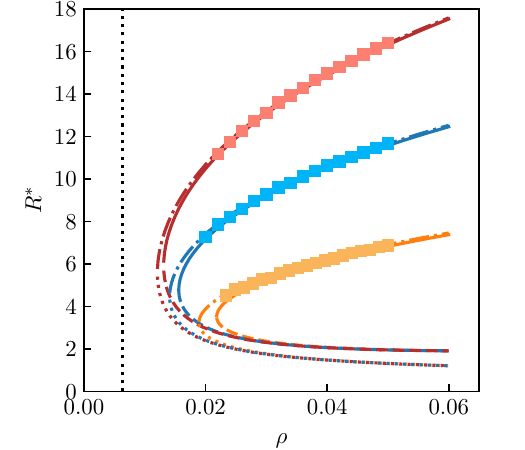}};
\node[] at (0.2,2.7) {(b)};
\node (fig3) at (-3,-5.5) {\includegraphics[width=0.45\textwidth]{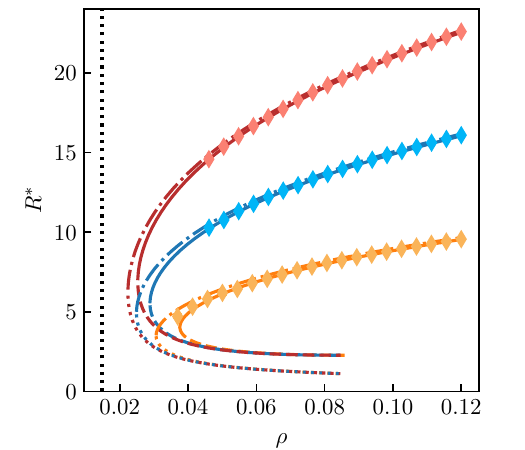}};
\node[] at (-6,-2.8) {(c)};
\node (fig4) at (3,-5.5) {\includegraphics[width=0.45\textwidth]{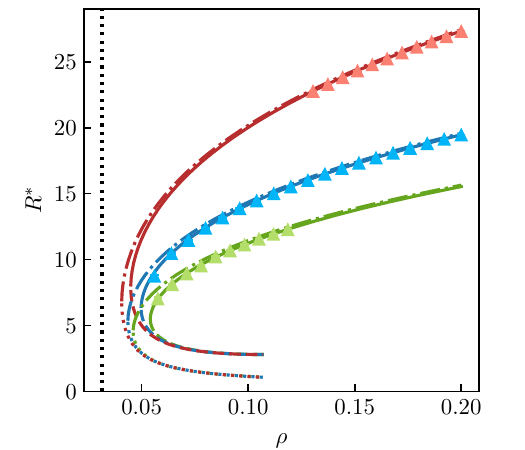}};
\node[] at (0.2,-2.8) {(d)};
\node (fig5) at (-3,-11) {\includegraphics[width=0.45\textwidth]{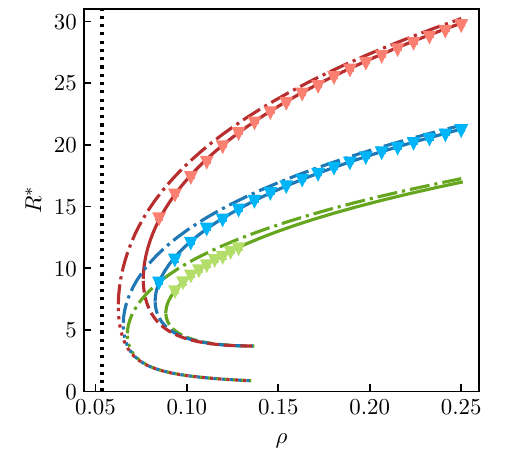}};
\node[] at (-6,-8.3) {(e)};
\node (fig6) at (3,-10.7) {\includegraphics[trim={1.7cm 10cm 21.5cm 1cm},clip, width=0.45\textwidth]{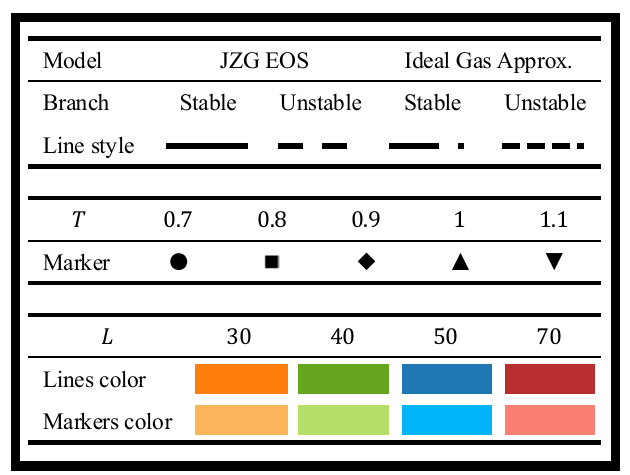}};
\end{tikzpicture}
\caption{Evolution of critical radius against box density of confined system with different box sizes and at different temperatures: (a) $T=0.7$, (b) $T=0.8$, (c) $T=0.9$, (d) $T=1$, and (e) $T=1.1$. Legends are in the bottom right. Color lines represent predictions and markers represent MD results. Vertical dotted black lines indicate the binodal limit for each temperature.}
\label{fig:ConfinedSystemSolutions}
\end{figure}
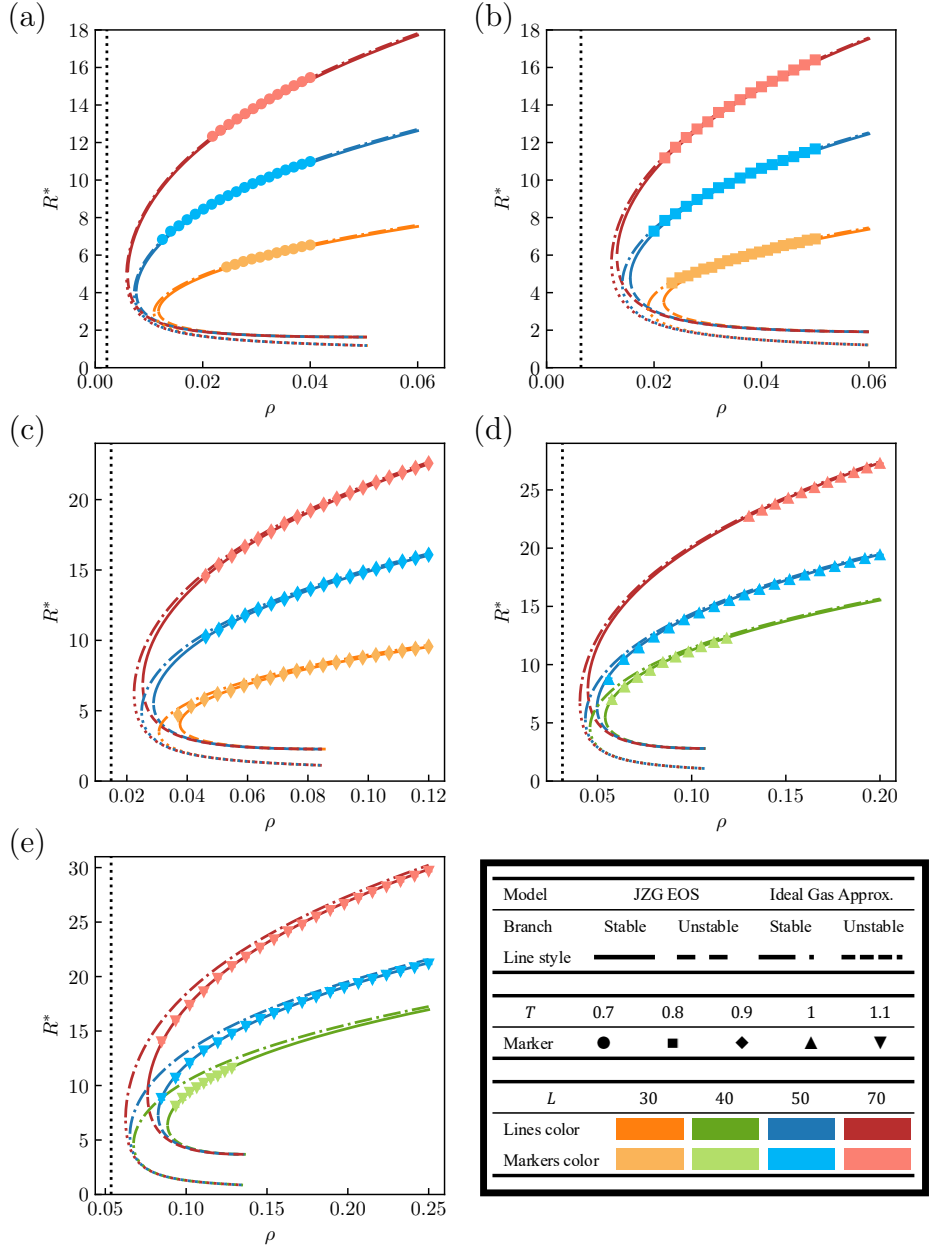

Fig.~\ref{fig:ConfinedSystemSolutions}(a) presents the two solution branches as a function of box density $\rho$ at $T=0.7$ for various box sizes, $L=30$ (red), $50$ (blue), and $70$ (orange), obtained using the JZG EOS and the computed surface tension. 

The smaller droplet (dashed lines) corresponds to the unstable equilibrium, while the larger droplet (solid lines) represents the stable cluster. As expected \cite{Reguera_2003,Wilhelmsen_2015,Philippe_PRE_2017}, the two branches merge for the density at which the superstabilization effect operates, i.e., no cluster forms since the vapor phase at this given box density becomes stable. For supersaturation between the binodal limit (vertical black dotted line) and this box density, nucleation is therefore impeded \cite{Reguera_2003}. 
Predictions obtained from the ideal gas approximation are also shown: stable and unstable branches are represented by dash-dotted and dotted lines, respectively. They agree very well with the JZG EOS results, particularly for the upper branches corresponding to the stable clusters.

The seeding technique is then employed to find the stable clusters at this temperature. For each box size, a single simulation run is performed. For example, in the case $L=70$ a liquid seed of density $\bar{\rho}_l$ is introduced with an initial radius $R$ chosen very close to the theoretical predictions for the highest box density $\rho=0.04$. Mass conservation then determines the vapor density of the initial state. An NVT simulation is run for a sufficiently long time (typically $5000$ to $10000$ $\tau$) to equilibrate the liquid cluster with the vapor. Equilibration is very fast, since the initial seed is prepared near the targeted cluster. The radius of the stable cluster, $R^*$, for this box size and density is averaged over the last 100 stable configurations (dumped every $5\tau$). 
To explore a lower box density, vapor particles are randomly removed, after which a new NVT run is carried out to equilibrate the cluster. This process is repeated iteratively to obtain stable liquid clusters at a range of box densities. The key advantage of this procedure is that the initial seed is always prepared close to the stable cluster, resulting in very efficient seeding simulations. 

In order to measure the cluster properties, we use \texttt{Ovito}~\cite{ovito} and first segment the liquid droplet by clustering analysis. A clustering distance, $d$, is defined, and particles forming a connected set --- each within a distance $d$ of at least one other --- are identified; the number of such particles are denoted $N_l^*$. The clustering analysis is not very sensitive to $d$ within the conventional range $[1.2,\,1.5]$ for Lennard-Jones droplets~\cite{Diemand_JCP2013}; in our analysis we adopt $d=1.25$. Several methods available in \texttt{Ovito}'s MD standard analysis library can then be used to determine the cluster radius: (i) fitting the radial density profile, or its integral, in simulation to a modified sigmoidal function~\cite{Diemand_JCP2014,Lutsko_2008}, (ii) the convex hull method \cite{ovito}, or (iii) the radius of gyration \cite{ovito}. The radius may also be estimated from the droplet volume, $R^* = (3N_l^*/(4\mathrm{\pi}\bar{\rho}_l))^{1/3}$, assuming the droplet is homogeneous with bulk liquid density $\bar{\rho}_l$ from the phase diagram. In our case, all methods yield very similar results, since the clusters are well defined, nearly spherical, and relatively large. For sake of clarity, only the results obtained from the droplet volume estimation are shown.
The corresponding results are also shown in Fig.~\ref{fig:ConfinedSystemSolutions}(a) as light-colored circles with error bars. The error bars, representing standard deviations in $R^*$, are too small to be visible. With three configurations ($L=30$, $50$, and $70$), stable clusters with radius ranging from $5$ to $16$ are obtained. Remarkably, the seeding results are reproduced very well by CNT predictions using either the JZG EOS or the ideal gas approximation. Fig.~\ref{fig:ConfinedSystemSolutions}(b)-(e) present seeded simulation results at higher temperatures: $T=0.8$, $T=0.9$, $T=1$, and $T=1.1$. Agreement with CNT predictions using the JZG EOS remains excellent across this range. In contrast, the ideal gas approximation becomes unreliable at high temperature. Nevertheless, it provides sufficiently accurate estimates to serve as practical initial conditions for seeded simulations, owing to the simplicity of the corresponding CNT calculations.

\begin{figure}[p]
\begin{tikzpicture}
\node (fig1) at (-3,0) {\includegraphics[width=0.45\textwidth]{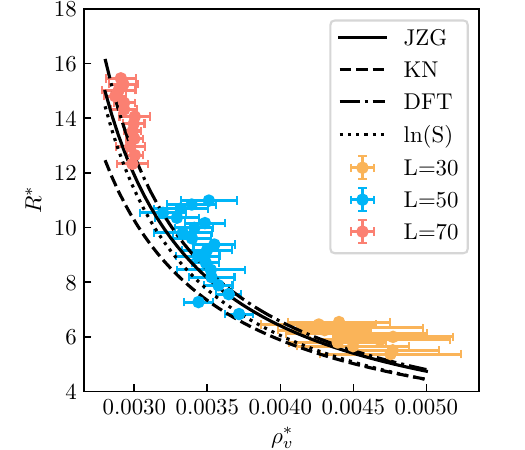}};
\node[] at (-6,2.7) {(a)};
\node (fig2) at (3,0) {\includegraphics[width=0.45\textwidth]{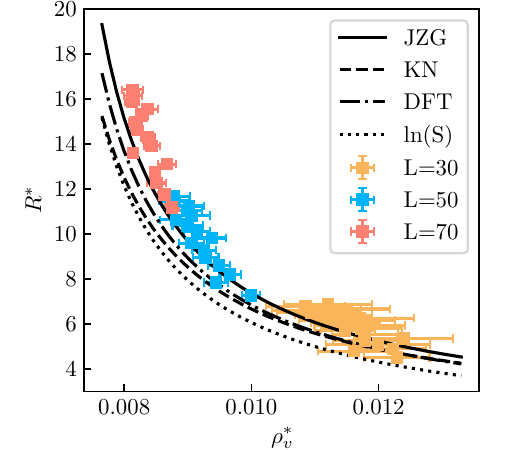}};
\node[] at (0.2,2.7) {(b)};
\node (fig3) at (-3,-5.5) {\includegraphics[width=0.45\textwidth]{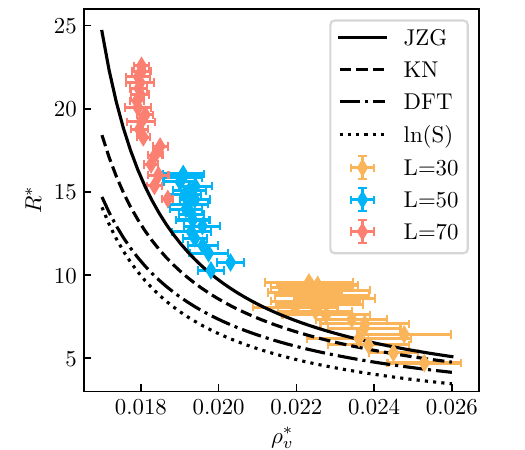}};
\node[] at (-6,-2.8) {(c)};
\node (fig4) at (3,-5.5) {\includegraphics[width=0.45\textwidth]{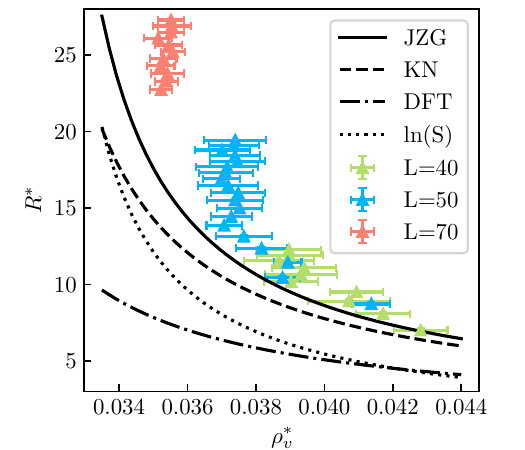}};
\node[] at (0.2,-2.8) {(d)};
\node (fig5) at (0,-11.5) {\includegraphics[width=0.45\textwidth]{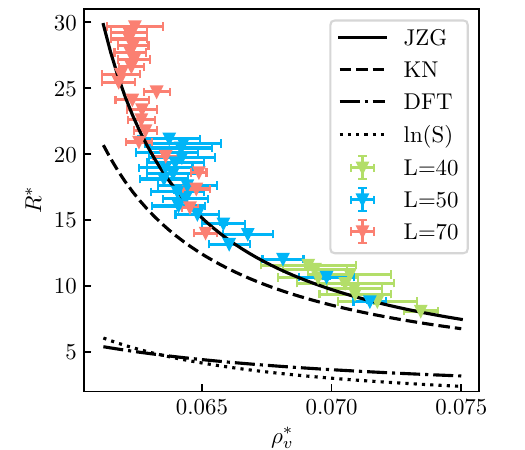}};
\node[] at (-3,-8.8) {(e)};
\end{tikzpicture}
\caption{Evolution of critical radius against vapor density at different temperatures: (a) $T=0.7$, (b) $T=0.8$, (c) $T=0.9$, (d) $T=1$, and (e) $T=1.1$. Markers represent measurements from seeded MD simulations of confined system, and their error bars represent standard deviations in $\rho_v^*$ and $R^*$. Black lines represent CNT predictions with different thermodynamic models.}
\label{fig:R_rhov}
\end{figure}
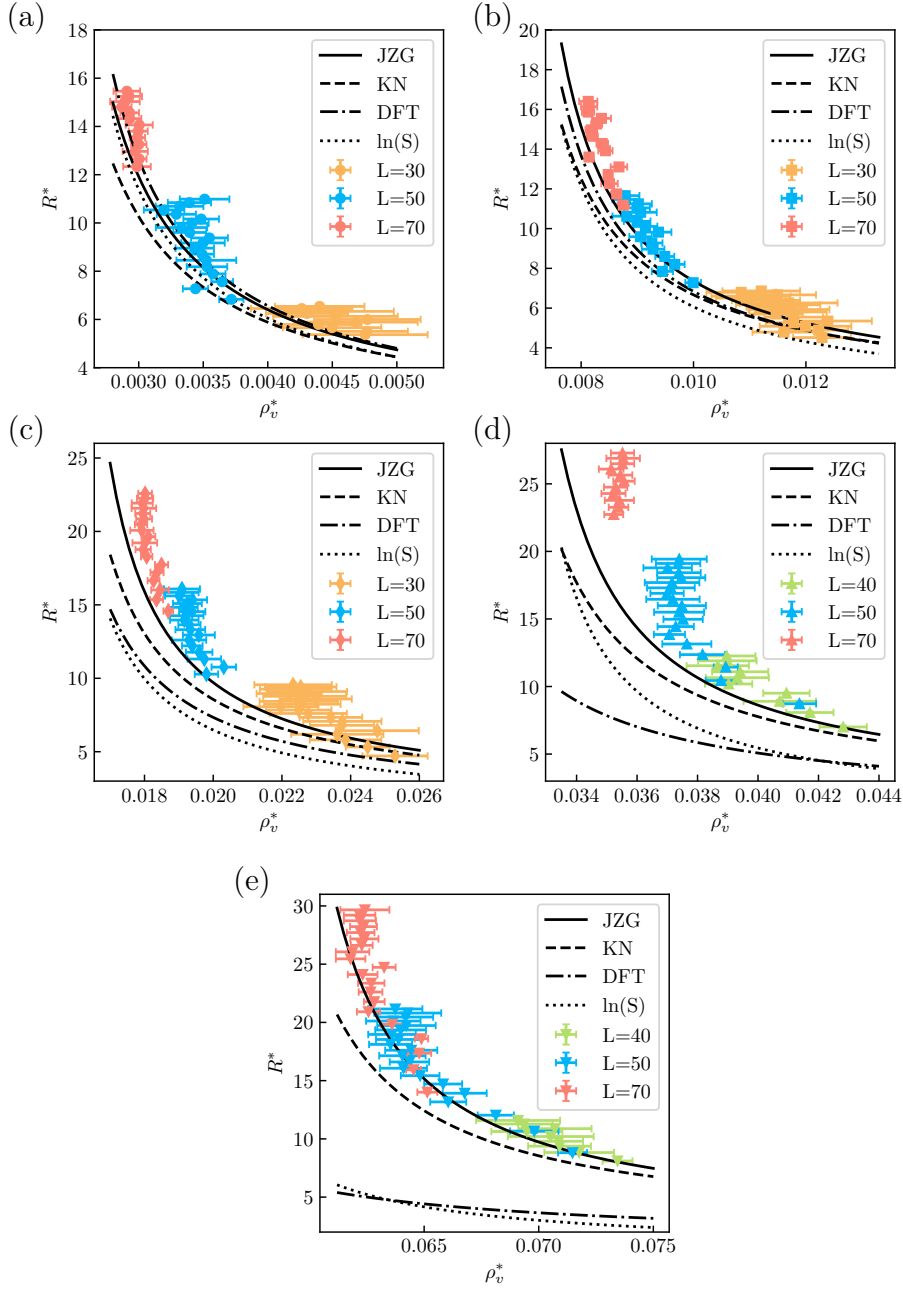

As mentioned earlier, the purpose of seeded simulations in the NVT ensemble is to achieve the equilibrated configuration corresponding to a stable cluster, characterized by radius $R^*$ and density $\rho_l^*$, in equilibrium with a vapor phase of density $\rho_v^*$. The same cluster, in fact, is unstable in the NPT and $\mu$PT ensembles, as well as in infinite systems at vapor density $\rho_v^*$~\cite{Vega_JCP2023,Wilhelmsen_2015}. In these cases, the cluster still satisfies the equilibrium conditions, Eqs.~(\ref{eqmu}) and (\ref{laplace}), but under conditions of fixed supersaturation. Nucleation is frequently studied under such circumstances, where either the pressure or the chemical potential of the surrounding phase is imposed. The equilibrium vapor density of a seeded simulation is estimated by $\rho^*_v = \left(N-N_l^*\right) / \left( V - 4\mathrm{\pi} R^{*3}/3 \right)$. The evolutions of critical cluster radius against vapor density at different temperatures are presented in Fig.~\ref{fig:R_rhov} by light-colored markers with error bars. 

The legend for markers is identical to that in Fig.~\ref{fig:ConfinedSystemSolutions}. The error bars represent standard deviations in $\rho_v^*$ and $R^*$, while the latter is invisible in the figure. At each temperature, the expected dependence of cluster size on supersaturation is recovered: the critical radius diverges as the vapor density approaches the binodal limit. As expected, fluctuations tend to increase with temperature and are more pronounced for smaller box sizes, primarily due to fluctuation in radius having a more substantial impact on vapor. CNT predictions by Eqs.~(\ref{eqmu}) and (\ref{laplace}) at fixed $\rho_v^*$ are also reported as black lines. Several thermodynamic models are employed to compute the chemical potential: the JZG EOS as previously, the Kolafa and Nezbeda (KN) EOS~\cite{kolafa_lennard-jones_1994}, classical density functional theory (DFT) with perturbation theory~\cite{lutsko_effect_2005}, and the ideal gas approximation for the vapor phase~\cite{Reguera_2003} as Eq.~(\ref{approxR}). This constitutes a stringent test of the accuracy of these models, since CNT is expected to provide reliable predictions at low supersaturation, the regime relevant for the seeding technique.
Overall, CNT predictions obtained with the JZG EOS show very good agreement with MD seeded simulations across the entire temperature range studied. The KN EOS also performs relatively well, especially at low temperature. Predictions based on DFT are accurate at low temperature but not so good at higher temperature, reflecting the limitations of DFT in describing bulk properties in this regime~\cite{peng_density_2008}. The ideal gas approximation also breaks down at high temperature, indicating that this approach should be employed with considerable caution.

\section{Summary}
\label{sec:conclusion}
The seeding technique has been applied to investigate the condensation of Lennard-Jones particles and, at the same time, to test both CNT and the accuracy of widely used thermodynamic models for Lennard-Jones fluid. The properties of the critical clusters predicted by CNT are found to agree well with seeded simulation results when JZG EOS is employed. The KN EOS also performs relatively well. In contrast, caution is required when using DFT or the ideal gas approximation, as both exhibit significant inaccuracies at high temperatures ($T \geq 0.9$). Nevertheless, the ideal gas approximation in confined systems, as expressed in Eqs.~(\ref{masscons}) and (\ref{approxR}), remains of very practical value. It provides a simple and effective framework for initializing and guiding NVT seeded simulations, especially when no thermodynamic model for chemical potential is available. We restrict ourselves to the gas-liquid phase transition but the seeding technique was proved to be also a powerful technique for studying crystal nucleation \cite{Vega_Seeding_JCP2016,Sun_PRL2018}. Coupled with CNT calculations, NVT seeded simulations are anticipated to be of great interest for quantitative investigation of materials provided realistic interatomic potentials exist.

\section*{Acknowledgments}
\normalsize This work was supported by the ANR TITANS project.

\end{document}